\documentclass{article}
\title{On Problem of Mass Origin and Self-Energy Divergence 
in Relativistic Mechanics and Gravitational Physics}
\author{A. Vankov}
\date{Bethany College, KS; vankova@bethanylb.edu; anatolivankov@hotmail.com}

\begin{document}
\maketitle

\begin{abstract}
The classical problem of self-energy divergence was studied in the framework 
of Lagrangian formulation of Relativistic Mechanics. The conclusion was made that 
a revision of mass-energy concept is needed for 
the development of singularity-free gravitational and electromagnetic 
field theory. Perspectives of the development of unified field theory are discussed.
\end{abstract}

\section{Introduction}
\label{section.Introduction}
The problem of mass is central in Gravitational and Particle Physics,
Astrophysics, Cosmology and field theories. In Kinematics of Special
Relativity Theory (SRT), the total mass $m_{tot}$ of a point-like particle
is related to the total energy by $E_{tot}=m_{tot}c^2$. A constant proper mass
$m_0$ and kinetic mass $m_{kin}$ are relativistic components of the
total mass:
\begin{eqnarray}
m_{tot}=\gamma m_0, & m_{kin}=(\gamma -1)m_0
\label{I.1}
\end{eqnarray}
where $\gamma=1/\sqrt{1-\beta^2}$ is the Lorentz factor for a relative speed
$\beta=u/c$ in a given inertial reference frame. As concerns dynamical
mass properties, we found that the proper mass depends on the potential of force
field, on the gravitational and the Coulomb potential, in particular.
Consequently, potential and kinetic energy become defined in SRT as strictly as
in Newtonian Physics but at a new (relativistic) level of understanding. 
This allowed us to gain an insight into the known self-energy divergence problem.
The problem is seemingly resolved in Quantum Electrodynamics which became 
a successful theory after an implementation of somewhat artificial
renormalization procedures. The idea was to get
rid of singularities by introducing an asymptotically hidden
cut-off parameter. However, GRT was shown to be nonrenormalizable, most likely, because 
of its principally new concepts of space-time and momentum-energy  
\cite{1, 2, 2a}.  
Ideally, an originally divergence-free theory rather than a normalizable one 
is desired. Our finding was that such a theory 
should follow from the Lagrangian formulation of Relativistic Mechanics: 
if consistently realized, SRT Mechanics  
leads to a variable proper mass. Consequently, the proper mass ``exhaustion'' under
strong-field conditions takes place, which results in the elimination of singularities 
in the Coulomb and gravitational potentials. 
But could one extend the SRT-based approach to the gravitational physics area? It is widely 
believed that SRT is incompatible with the gravity phenomenon; numerous attempts to
incorporate it into SRT failed (see, for example,
\cite{2}). Main ``incompatibility'' arguments  
have arisen from experimental tests involving gravitational properties of light.
The results were treated 
in terms of the GRT metric and the constant proper mass concept, 
which we believe have to be revised. Further, it is shown that the 
SRT-based alternative approach provides 
as adequate description of gravitational \em weak-field \em
experiments as GRT does but with a different physical interpretation. 
At the same time, predictions of \em strong-field \em
effects were found to be drastically different. In particular, 
we predict an existence of superluminal
particles in a gravitational field, and propose a corresponding
experimental test involving cosmic ray ultra-high energy particles 
in the gravitational field of Earth.

The purpose of this work is to present results of
our study of the classical problem of self-energy divergence in the framework 
of Lagrangian formulation of Relativistic Mechanics. The conclusion is made that
the development of the SRT-based divergence-free gravitation and electromagnetic 
field theory is possible. Speculations on a development of unified quantum 
field theory are discussed.
\section{On physical principles of SRT Mechanics with gravitational forces}
\label{section.1}
\subsection{A relativistic mass-energy concept and a proper mass variation}
\label{section.1.1}
In the Lagrangian formulation of Relativistic Mechanics of a single particle, 
the rate of 4-momentum change equals the Minkowski force. A
variation of the proper mass follows from the corresponding SRT dynamical
equations (using Synge's denotations \cite{3}):
\begin{eqnarray}
\frac{d}{ds}[m(s)\frac{dx_i}{ds}]=K_i & (i=1,2,3,4)
\label{1.1}
\end{eqnarray}
They describe a particle motion on a world line $x_i(s),
\vec{x}=\{x_1,x_2,x_3,ict\}$, with a 4-velocity $dx_i/ds$, 
where $K_i(s)$ is a Minkowski 4-force vector, and $s$ is a line arc-length.
By definition of a time-like world line of a massive particle, we have the
fifth equation:
\begin{equation}
\sum_i\frac{dx_i}{ds}\frac{dx_i}{ds}=-1
\label{1.2}
\end{equation}
that makes the problem definite with respect to five unknown functions:
$x_i(s)$ and $m(s)$.
The proper mass variation along the world line is explicitly seen from the
next equation obtained from (\ref{1.1}) and (\ref{1.2}):
\begin{equation}
\frac{dm}{ds}\frac{dx_i}{ds}+m\frac{d^2x_i}{ds^2}=K_i
\label{1.3}
\end{equation}
For the sake of convenience, one may consider the description of motion in 
3-space $(\alpha=1, 2, 3)$ and time $t$ $(i=4)$ rather than in
spacetime $(i=1, 2, 3, 4)$ using the relation $ds=cdt/\gamma$ and
formulas for relative (``ordinary'') forces $F_\alpha$:
\begin{equation}
F_\alpha=c^2K_\alpha/\gamma
\label{1.4}
\end{equation}
Now the equations of motion take the form:
\begin{equation}
\frac{d}{dt}(m\gamma u_\alpha)=F_\alpha
\label{1.5}
\end{equation}
\begin{equation}
c^2\frac{d}{dt}(\gamma m)={\bf F}\cdot {\bf u}+\frac{c^2}{\gamma}\frac{dm}{dt}
\label{1.6}
\end{equation}
where $u_\alpha (t)=dx_\alpha /dt$ $ (\alpha=1,2,3)$ is the 3-velocity, and the proper
mass $m$ is dependent on  space and time coordinates in a given inertial
reference frame. On the right-hand side of (\ref{1.6}) the term
$(\frac{c^2}{\gamma}\frac{dm}{dt})$ is recovered to account for the proper
mass variation in a force field. The effect of the proper mass variation 
was noted in \cite{3,4} but was never paid attention in literature. For example, 
the fact that a paticle speed cannot exceed the speed of light is often 
illustrared by the expression of motion of the partile   
driven by a constant inertial force $f_0=Const$ : 
\begin{equation}
cp(t)=f_0 t, \  \beta(t)={f_0t}/{\sqrt{m^2c^2+f_0^2 t^2}}
\label{1.7}
\end{equation}
where the momentum $p(t)$ is proportional to the time elapsed. The mass in (\ref{1.7}) 
is supposed to be a constant proper mass $m_0$. To check it, one has to
consider a general problem on acceleration of the particle by a pulse of force
with transients specified. It follows from (\ref{1.5}) and (\ref{1.6}) that 
the proper mass varies during trantients. When the force reaches a plateau it becomes 
constant but different from $m_0$, the difference being 
a binding energy of a particle in the
system ``particle-accelerator''. In the end of the pulse the proper mass
acquires the initial value $m_0$ in a new state of free motion with kinetic
mass-energy (\ref{I.1}) taken from the accelerator. \em A dynamical change
of the proper mass is a manifestation of a potential difference developed
between the particle and the accelerator; consequently, the interaction should be 
characterized by the corresponding mass-energy current. \em  
A general relativistic mass-energy formula following from
(\ref{1.5}) and (\ref{1.6}) holds:
\begin{equation}
E(t)^2=p(t)^2c^2+m(t)^2c^4
\label{1.8}
\end{equation}
It describes the instantaneous state of a single particle in a force field and leads to
(\ref{1.7}) under a constant force condition. For a free motion, the equation
(\ref{1.8}) is reduced to (\ref{I.1}) and the known SRT Kinematics formula
\begin{equation}
E^2=p_0^2c^2+m_0^2c^4=Const
\label{1.8a}
\end{equation}
We shall see further that for a particle in free fall in a static
gravitational field the expression (\ref{1.8}) takes the form of the total
energy conservation law with the proper mass being variable
\begin{equation}
E^2=p(t)^2c^2+m(t)^2c^4=m_0^2c^4
\label{1.8b}
\end{equation}

\subsection{A relativistic generalization of a gravitational potential 
and an elimination of singularities}
\label{section.1.3}

To understand a physical meaning of equation (\ref{1.8b}), let
us consider a point-like particle of proper mass $m_0$ 
in a spherical symmetric gravitational field; the latter is characterized by 
a classical potential $\phi(r)$ due to a uniform sphere of mass  $M\gg m_0$ and a radius $R$:
\begin{eqnarray}
\phi(r)=-c^2(r_g/r), & r\ge R
\label{1.9}
\end{eqnarray}
where $r_g=GM/c^2$ is a ``gravitational radius'' ($G$ is the universal 
gravitational
constant), $r$ is a distance from the center of the sphere.
So far, we assume that $R\ge r_g$.
The potential (\ref{1.9}) is defined per unit mass which could be 
a rest mass of a test particle in the Newtonian Mechanics. 
In SRT Mechanics the proper mass $m$ must be field dependent. Imagine that the particle 
can be slowly moved with a constant
speed along the radial direction with the use of an ideal
transporting device supplied with a recuperating battery. Thus,
the particle will exchange energy with the battery in a process of
mass-energy transformation prescribed by the SRT mass-energy concept. The
change of potential energy of the particle is related to the change
of the proper mass:
\begin{eqnarray}
dm=-m(r)d(r_g/r), & r\ge R
\label{1.10}
\end{eqnarray}
Thus, the proper mass of the particle is a function of the distance r:
\begin{eqnarray}
m(r)=m_0\exp(-r_g/r), & r\ge R
\label{1.11}
\end{eqnarray}
where $m_0$ is a proper mass at infinity. In a weak field
approximation $(r\gg R)$ we have
\begin{equation}
m(r)\cong m_0(1-r_g/r),
\label{1.12}
\end{equation}
with a Newtonian limit $m(r)=m_0$ at $r_g<<R$. At
$(r_g/r)\to \infty$ the proper mass tends to exhaust.       

Once the proper mass variation is taken into account, a gravitational 
force takes a kinematical form:
\begin{equation}
F(r)=m_0c^2(r_g/r^2)\exp (-r_g/r)
\label{1.13}
\end{equation}
The same result follows from (\ref{1.5}) and (\ref{1.6})
when the interaction of the particle with the
battery is taken into account. 
One can find a relativistic generalization of the static potential function
(\ref{1.9}): 
\begin{eqnarray}
m_0\phi(r)=\int\limits_r^{\infty} F(r,m(r))dr=-m_0c^2[1-\exp(-r_g/r)], &
r\ge R
\label{1.16}
\end{eqnarray}
The expressions (\ref{1.13}) and (\ref{1.16}) have a point-like particle limit. 
In geeneral, the proper mass of a test particle at a point
$\bf r$ in 3-space uniquely characterizes a static gravitational field $\phi({\bf r})$:
\begin{equation}
m({\bf r})/m_0=[1+\frac{1}{c^2}\phi({\bf r})]
\label{1.17}
\end{equation}
The potential changes within the range $-c^2\le \phi(r)\le 0$; therefore, it
is limited by the factor $c^2$. This is a result of fundamental
importance. It shows that \em a singularity is absent 
in the relativistic form of gravitational potential\em.

\subsection{A source of the Coulomb potential}
\label{section.1.4}

The Coulomb potential exhibits a similar effect of ``exhaustion'' of the
proper mass. Let us consider two electrically interacting particles with
equal proper mass $m_0$ and unlike electric charges of an equal magnitude
$q$. With the use of ``ideal transporting devices'', we can move them
uniformly along an $x$-axis, the center of mass being fixed.  One
can find a proper mass dependence on the potential:
\begin{eqnarray}
m(x)=m_0(1-x_a/x), & x\ge x_a
\label{1.18}
\end{eqnarray}
where ``the annihilation parameter'' is $x_a=k_0q^2/2m_0c^2$ ($k_0$ is the
electric constant at infinity). At $x=x_a$ the proper mass vanishes (the
particles annihilate), when the force reaches a maximal value and vanishes
at smaller distances. An expression for a proper mass variation for a
repulsive force may be found by replacing the ``minus ''sign''  by a ``plus'' in
(\ref{1.18}). When the inertial force ``pushes'' the particle towards the
repulsive center, the battery spends energy, and the proper mass
increases indefinitely: $m(x)\ge m_0$. 

If the particles interact gravitationally, this model gives the result: 
\begin{equation}
m(x)=m_0/(1+x_g/x)
\label{1.19}
\end{equation}
whith $x_g=Gm_0^2/2m_0c^2$ . The
gravitational force is many orders weaker than the electric one at $x>x_a$
but its action extends to the extreme point $x=0$; consequently, 
the range of a potential energy change per particle is the same
as for the electric attractive force: $-m_0c^2\le m\phi \le 0$. One can
verify the equality of total work $\int_0^\infty F_g(x)dx=\int_{x_a}^\infty
F_e(x)dx=2m_0c^2$ using parameterized relativistic expressions for
the static forces in the above examples: 
\begin{eqnarray}
F_g(x)=2m_0c^2\frac{x_g}{(x+x_g)^2}\nonumber\\
F_g(x)dx=-m_0c^2d(\frac{x_g}{x+x_g}),& 0<x<\infty\
\label{1.20}
\end{eqnarray}
\begin{eqnarray}
F_e=2m_0c^2\frac{x_a}{x^2}\nonumber\\
F_e(x)dx=-m_0c^2d(\frac{x_a}{x}),& x_a\le x\le\infty\
\label{1.21}
\end{eqnarray}
where $x_g$ and $x_a$ play the role of ``cut-off'' parameters in a classical
$1/x$ potential. Thus, a proper mass variation in a conservative field leads to a
natural elimination of the classical field singularities.  

The above results show that the proper mass plays the role of a common source
for both the gravitational and electric potential. Therefore, the electric force could be 
influenced by a sufficiently strong external gravitational field. In our imaginary
experiment, let us put a pair of charged particles on some equipotential surface in
the central gravitational field (\ref{1.9}) at $r\gg r_g$, a distance between particles
being  $x\gg\theta r$ ($\theta$ is a central angle subtending an arc
of length $x$). We expect
that under stated conditions an external gravitational field does not change
the gravitation-to-electric force ratio $F_g(x)/F_e(x)$. Then, we have a
common exponential factor in expressions for forces:
\begin{equation}
F_g(r)=\frac{Gm_0^2}{\theta^2r^2}\exp (-2r_g/r);\ 
F_e(x)=\frac{k_0q^2}{\theta^2r^2}\exp (-2r_g/r)
\label{1.22}
\end{equation}
This means that the gravitational field could influence the permittivity and
the permeability of space (according to observations, an electric charge is
not affected). The expression for the electric
force on a test charged particle of a mass $m_0$ and a charge $q$ in the
Coulomb (attractive) static field due to a charge $Q$ uniformly distributed
over a massive (not polarizing) sphere of a mass $M\gg m_0$ should include a
field dependent electric ``constant'' $k(r)$. The latter becomes proportional to the
proper mass: $m(r)=m_0\exp (-r_a/r)$, $k(r)=k_0\exp (-r_a/r)$, $(r\ge R)$,
where $r_a=k_0Qq/m_0c^2$ ($k_0$ is the electric constant at infinity). 
Thus, we have similar expressions for a gravitational and an attractive electric force:
\begin{eqnarray}
F_g(r)dr=-m_0c^2\exp (-r_g/r)d(\frac{r_g}{r}), & 0<r\le\infty
\label{1.22a}
\end{eqnarray}
\begin{eqnarray}
F_e(r)dr=-m_0c^2\exp (-r_a/r)d(\frac{r_a}{r}), & x_a<r\le\infty
\label{1.22b}
\end{eqnarray}
with parameters $r_g$ and $r_a$ characterizing field strength. 
They indicate that particles actually have a size. The gravitational radius relates to a volume,
which effectively comprises the particle proper mass-energy $m_0c^2$. Because the 
radius of electromagnetic interaction is many orders bigger, the gravitational 
force does not play any role in particle interactions. However, 
it is important in astronomical scales.   

\subsection{On Dynamics of a massive particle and a photon in a gravitational field}
\label{section.1.5}
Conservative field properties are embedded in
equations (\ref{1.5}) and (\ref{1.6}). Consequently, 
for a particle in free fall in the spherical symmetric
gravitational field (\ref{1.9}) a total mass is constant: 
\begin{equation}
m_{tot}=\gamma_rm(r)=m_0
\label{1.23}
\end{equation}
Putting the expression for a gravitational force
$F(r)dr=c^2m(r)d(r_g/r)$ into equations (\ref{1.5}) and (\ref{1.6}), 
we have for $r\ge R>r_g$:
\begin{equation}
m_0^2=m_0^2\beta(r)^2+m(r)^2
\label{1.24}
\end{equation}
\begin{equation}
m(r)/m_0=1/\gamma_r=(1-r_g/r)=\sqrt{1-\beta(r)^2}
\label{1.25}
\end{equation}
\begin{equation}
\beta(r)=u(r)/c=\frac{1}{c}\frac{dr}{dt}=\sqrt{1-(1-r_g/r)^2}
\label{1.26}
\end{equation}
The total energy conservation law is given in (\ref{1.24}) as a relativistic 
relationship between a varying proper mass and a momentum. The
expression (\ref{1.25}) shows that a kinetic energy gain is equal to 
the corresponding potential energy change. It is worth noting that in
these dynamical relations the $\gamma_r$ factor looks like a linear
approximation of the corresponding exponential factor in 
kinematic expression (\ref{1.12}); this is because
the equations of motion account for relativistic rescaling of space-time
coordinates under dynamical conditions, when the gravitational force
acquires Minkowski force properties. Finally, the expression (\ref{1.26})
describes a radial speed of a particle falling from rest at infinity. 
If the particle has an initial radial momentum
$\gamma_0\beta_0m_0c$ ($\gamma_0>1$), then, taking into account the total mass conservation
$\gamma_0^2m_0^2=\gamma_0^2m_0^2\beta(r)^2+m(r)^2$ and the mass dependence on field
$m(r)/m_0=1-r_g/r$, we have $\gamma=\gamma_0\gamma_r$, and the expression (\ref{1.26}) is modified:
\begin{equation}
\beta(r)=\sqrt{1-(1-r_g/r)^2/\gamma_0^2}
\label{1.27}
\end{equation}
The solution formally shows
that the proper mass vanishes at $r=r_g$. Because a baryon charge of a
single particle cannot be destroyed, we have to conclude that the above
case cannot be physically realized: the results are valid at $r\ge R>r_g$. 
They show that a particle carrying a non-zero proper mass 
in free fall can never reach the ultimate
speed of light, though it constantly accelerates (the condition
$\beta(r)<1$, $d\beta/dr>0$ always takes place).


Next, let us consider a radial motion of a photon in gravitational field.
Unlike the particle, the photon does not have a proper mass. From 
equations (\ref{1.5}) and (\ref{1.6}) one can note that any force changes a
momentum through the action on a total mass. Because the total mass is
constant, the only way the photon can change the momentum is by
changing the speed. In other words, the speed should be influenced by the
potential $\phi(r)$. 
The following expression is consistent with SRT
Mechanics:
\begin{equation}
\beta_{ph}(r)=c(r)/c_0=1-r_g/r
\label{1.28}
\end{equation}
Actually, this is the relative speed of wave propagation $c(r)=\lambda(r)$
with a constant frequency $f=Const$. The speed is constant on an
equipotential surface $r=r_0$; in this case, it may be termed a tangential,
or arc speed. Henceforth the speed of light at
infinity will  be denoted $c_0$. In addition to (\ref{1.28}), one can
define the radial ``coordinate'' speed $c^*(r)=\beta^*(r)c_0$. It is measured by the differential
time-of-flight method by an observer at infinity with the use of the so-called
standard clock. If a unit length were found from circumference
measurements, the radial scale would be determined by the field-dependent length
unit proportional to the wavelength of the standard photon emitted from
infinity $dr/d\lambda=(1-r_g/r)$. Therefore: 
\begin{equation}
\beta^*(r)=\beta(r)\frac{dr}{d\lambda}=(1-r_g/r)^2
\label{1.29}
\end{equation}
As is seen, the photon while approaching the sphere slows down and tends to
stop at $r\to R\to r_g<R$. Our analysis of the phenomenon led us to the conclusion 
that the photon propagates in space of a gravitational field as in a refracting medium.

The variation of the proper mass and the speed of light in a gravitational field is a consequence
of the SRT mass-energy concept. Both phenomena should be considered as a result of interaction 
of the particle or the photon with the field; they are crucial for a metric determination 
in Relativistic Mechanics and should be verified in experiments.

\section{Experimental basis of gravitational theory}
\label{section.2}
\subsection{Basic physical units and a metric determination}
\label{section.2.1}

In the relativistic world basic physical units became dependent on the 
gravitational field. How to measure them is the issue of theory foundations. Obviously, there are no
absolute measuring units: what we have from observations 
are ratios with respect to the
asymptotic values ``at infinity''. We consider the problem of metric determination in a broader 
sense than measuring the space-time metric alone. The complete metric 
should include also units of 4-momentum vector components along with 
the speed of light, all of them are to be reanalyzed in a framework of the Lagrangian formulation of 
Relativistic Mechanics. (Further, we assume that an electric charge is field-independent).
We see a solution of the problem in giving the proper mass a natural degree of
freedom. In this new mass-energy concept, a test particle and a photon 
reveal new gravitational properties of fundamental importance. First of all, 
there is ``an exhaustion'' of the proper mass and, correspondingly, the potential 
energy in a strong field; this effect eliminates field singularities. Secondary, 
an interaction of the photon with the gravitational field is different from what 
is assumed in current theories. The photon behaves in a gravitational field as 
in an optically active (refractive) medium rather than like a particle in the field. 
It means that there is no coupling of the photon to the gravitational field. 
Consequently, a probing the field with the test particle and the photon 
results in a new field characterization and new metric relations of physical quantities.

In further metric analysis we are going to show how  
to determine basic physical units
which could be transportable or reproducible throughout the space to make it
possible to compare their values in a field and ``at infinity''. Let us start with
choosing a standard (stable) test particle playing the role of a
quantum-mechanical oscillator when being used as a resonance emitter or a detector. 
A ``standard atomic clock'' will be the equivalent term for the standard
particle. Its variable (field-dependent) proper mass can be taken as a
variable unit of the proper mass. Then, characteristics of the electromagnetic
wave emitted by the standard clock may be considered in choosing the
standard units of length and time. From previous results, one can see that
the frequency $f_0$ of the photon \em emitted from infinity \em is a
field-independent standard quantity; therefore, the corresponding period $\delta t_0=T_0=1/f_0$
is a field-independent standard unit of time interval. We assert that the
proper resonance frequency $f_{res}$ of the standard atomic clock
in the field is proportional to the proper mass at the radial point $r'$, 
as in (\ref{1.25}) for a spherical
symmetric field. One may notice that the radial factor
$\gamma_r=1/(1-r_g/r')$ plays a gauge role in our metric relations. Thus, the
frequency depends on a radial position of the atomic clock in the field, as
next: $f_{res}(r')=f_{res}^0(1-r_g/r')\propto m_0(1-r_g/r')$. By definition,
the inverse quantity is the field-dependent proper time interval
$\delta\tau$ of the standard atomic clock at point $r'$:
$\delta\tau(r')=1/f_{res}(r')=\delta\tau_0/(1-r_g/r')$, where $\delta\tau_0$
is the proper time interval at infinity. It should be noted that so far we
do not differentiate between the electromagnetic wave of light and the
photon because the radial dependence of speed (\ref{1.28}) is assumed to be
the same for all frequencies (there is no dispersion). We admit that at
ultra-high energy this assumption may be not valid.

Next, let us look for a field-independent unit of length. The instantaneous
proper wavelength of the photon at any emission point $r'$ can play 
this role. The wavelength is a ratio of the speed of wave propagation
(\ref{1.28}) and the resonance frequency of an emitter $f_{res}(r')$;
therefore, the standard \em emission \em wavelength $\lambda_0$ is constant
and reproducible everywhere. We come to the important conclusion that \em
space-time mapping is possible in terms of field-independent units\em. In
general, the following formulas describe the field dependent proper time
interval of the standard atomic clock and characteristics of the photon at 
point $r$, if emitted by the standard atomic clock at point $r'$ (both the
emitter and the detector being at rest in a sphere centered reference frame): 
\begin{equation}
\delta\tau(r')=1/f_{res}(r')=\delta\tau_0/(1-r_g/r')
\label{2.1}
\end{equation}
\begin{equation}
f_{ph}(r'\to r)=f_0(1-r_g/r')
\label{2.2}
\end{equation}
\begin{equation}
c_{ph}(r'\to r)=c_0(1-r_g/r)
\label{2.3}
\end{equation}
\begin{equation}
\lambda_{ph}(r'\to r)=\lambda_0\frac{(1-r_g/r)}{(1-r_g/r')}
\label{2.4}
\end{equation}
It is easy to incorporate in these formulas the Doppler effect caused by a
relative motion of an emitter and a detector. The formula
(\ref{2.1}) describes the gravitational time dilation: caused by a proper mass
dependence on potential (\ref{1.25}). From (\ref{2.1}) and (\ref{2.2}),
it follows that there is a shift between the standard resonance frequency of
the emission line at point $r'$ and the absorption (detection) line at the
point $r$ (taking into account that the photon does not change the frequency
in flight). As is seen from (\ref{2.3}), the speed of the photon is a
function of a radial position; it does not depend on the location of the
emitter (see also (\ref{1.28})). From (\ref{2.4}), the wavelength of the
photon at the moment of emission at $r=r'$ is field independent and equals to
$\lambda_0$ (transportable standard length unit). Notice that the proper
time of atomic clock at some point at rest is measured as the oscillation
period of the photon at the point of emission, and the photon carries this
number during its flight over space filled with a gravitational field.  

In above, we did not need, in principle, (and for this reason did not
mention) a standard rod as a transportable equivalent of
$\lambda_0$, though it plays an important role in the GRT metric philosophy. 
This is a controversial issue because the absolutely rigid body should have 
infinite internal energy.  By saying this, we
do not discard standard rods from some specific imaginary experiments, if justified. 
For example, one may admit a local reproduction of standard measuring rods
($\lambda_0$-duplicates) for using them within a thin equipotential shell
without transporting a locally made piece into other shells. With this
remark, we agree, in principle, with the bookkeeper philosophy of ``rigid''
environment in imaginary gravitational experiments \cite{1}, but our
physical interpretation of measuring procedure will be different.

In the list of metric relations the speed of light plays a special
role. The arc speed and the coordinate speed are different due to field
medium anisotropy; they are the same in isotropic medium (a
uniform field). As was noted, a field acts on a photon as a refractive 
medium with the index of refraction $n=1/\gamma_r=(1-r_g/r)$, 
which plays a role of a gravitational gauge factor.
We have to conclude our metric analysis with the statement that 
the metric concept under discussion is introduced consistently with 
principles of Relativistic Mechanics, and it is falsifiable
in the following sense: a) the metric relations (\ref{1.24}-\ref{2.4}), 
and their uniqueness can be
experimentally tested and used for space-time mapping and mass-energy
scaling in the \em Minkowski space framework \em; b) integral experimental data related to 
gravitational properties of particles and photons can be treated in 
Relativistic Mechanics (as discussed further).

\subsubsection{Proper time interval and conservative field properties}
\label{section.2.1.2}

As was emphasized, SRT Mechanics describes a point-like particle motion on a
\em world line ${\bf x}(s)$ in Minkowski space\em. Having found a solution
to any particular problem, one is able, if interested, to calculate the
Minkowski metric form $ds^2=c_0^2dt^2-(dx^2+dy^2+dz^2)$,
where $x$, $y$, $z$ are usual Euclidian coordinates. The infinitesimal arc
length $ds$ is the time-like interval (a distance between two
neighboring points). Consequently, $ds=c_0d\tau=c_0dt/\gamma$, 
where $\tau$ is the proper time of the particle (the atomic
clock). It is constant everywhere (Lorentz invariant) in the particular
case $\gamma=Const$ of the test particle moving by inertia. That means that
the trajectory of a particle in ``free space'' is a straight line, and the
symmetry of translations and Lorentz transformations manifests a 4-space being
the Minkowski space free of sources. If a field is introduced, the world
line deviates from the straight line, and the proper time interval changes. 
Note that the introduction of the field is not associated with 
the inner curvature of space: the curved space concept would be a foreign body 
in the theory. Now, we are interested in determining an instantaneous
proper time interval of a test particle freely floating in the gravitational field 
in the coordinate system of an observer resting at infinity. 
Recall that the particle is considered the standard atomic clock. The observer can find it 
by measuring an instantaneous improper time interval and comparing it with the rate 
of his standard clock (similarly to imaginary experiments in SRT). 
The interpretation of the comparison procedure follows from our mass-energy concept: 
the proper time interval is the measure of the proper mass of the particle while the improper 
time interval is the Lorentz factor $\gamma=\sqrt{1-\beta^2}$ bigger, where 
the relative speed is meant with respect to the rest observer at infinity. Thus, the improper 
time is constant due to the total mass-energy conservation. 
A bound standard clock (attached to a shell or in orbital motion) has a smaller 
amount of total energy as compared to the similar clock in a hyperbolic motion. For example, 
the improper time of a particle in the shell of radius $r$ is $dt=d\tau_0/\gamma_r$. 
For a free motion at infinity, the SRT relations are $m_{tot}=\gamma_0 m_0$ and $dt=\gamma_0 d\tau_0$; 
consequently, the metric form for a free floating state of the particle with $\gamma_0>1$ is     
\begin{equation}
ds(r)=c_0d\tau(r)=c_0dt(r)/\gamma
\label{2.6}
\end{equation}
with the kinematical Lorentz factor 
$\gamma=1/\sqrt{1-\beta(r)^2}=\gamma_0\gamma_r$, as in (\ref{1.27}), and 
the dynamical factor $\gamma_r=1/(1-r_g/r)$.
Therefore, relations between the field-dependent metric form $ds$, the proper time 
of atomic clocks rested at
infinity $d\tau_0$ and in a free-floating state $d\tau(r)$ are:
\begin{equation}
ds(r)=c_0d\tau_0/\gamma_r
\label{2.8}
\end{equation}
\begin{equation}
d\tau(r)=d\tau_0/\gamma_r
\label{2.8a}
\end{equation} 
In the particular case of the particle in free fall from rest at infinity  
we have $\gamma_0=1$, $\gamma_rm(r)=m_0$, and $dt=d\tau_0$; 
it describes conservation properties of gravitational field 
with the following identities characterizing      
a rotational symmetry in 4-coordinate and
4-momentum space:
\begin{eqnarray}
p^2/m_0^2c_0^2+m^2/m_0^2=1, & m/m_0=1/\gamma_r
\label{2.9}
\end{eqnarray}
\begin{eqnarray}
dr^2/d\tau_0^2c_0^2+d\tau^2/d\tau_0^2=1 , & d\tau/d\tau_0=1/\gamma_r=\sqrt{1-\beta^2}
\label{2.10}
\end{eqnarray}
The corresponding metric is $ds^2=c_0^2 d\tau_0^2-dr^2$; 
the angle and the rate of the Minkowski space rotation can be easily found. 
Having a solution to equations of
motion (\ref{1.1}) for some specific problem, one can find scalar products
of 4-vectors $s^2={\bf x}\cdot{\bf x}$, $p^2={\bf p}\cdot{\bf p}$ and
construct the antisymmetrical tensor $M_{ik}=x_ip_k-x_kp_i$ to check the
angular momentum conservation as well. \em The ``gamma'' transformation 
$q(r)\to q_0/\gamma_r)$ of basic metric units in Minkowski space
reflects the rotational symmetry being identical in the 4-coordinate 
and 4-momentum space and their inner connection. \em 
Obviously, this is the reflection of conservation field properties as well.

\subsubsection{Metric comparison}
\label{section.2.1.3)}

The above symmetry does not
appear in GRT because the latter has different physical foundations,  
the curved space-time metric being part of it. In fact, it was Albert
Einstein and his colleagues who first attempted to solve the gravitational
problem by finding an adequate description of gravitational
properties of matter in the Minkowski framework \cite{5}. 
The basic reason for the historical departure
from SRT was the alleged inability of SRT Mechanics to account for the observed
effect of the bending of light. The photon was considered a particle 
with a total (kinetic) mass $hf/c_0^2$
being subject to gravitation; 
thereafter, the photon was assumed to behave in a gravitational field
similarly to a fast moving \em massive \em particle. There was a
long-standing search for SRT-based gravitational theory compatible with
this photon concept: the photon-to-gravity coupling was thought to be a
proven physical reality (and so it is still thought). Along with ``the photon
bending effect'', the observed ``reduced circumference effect'' is often
presented in literature as a typical illustration of a space-time \em
intrinsic \em curvature
characterized by the metric tensor $g_{ik}$ with the metric form
$ds^2=\sum_{i,k}g_{ik}dx^idx^k$. GRT was considered a relativistic generalization
of the Newtonian field theory; a classical picture
of a ``photon attracted by a gravitationally force'' was replaced in GRT by a 
picture of a geodesic motion of the photon in a curved 4-space. At the same time, 
forces, potentials and other physical attributes of classical physics were 
liquidated. This was an introduction of radically 
``new physics'', which eventually put GRT in an isolation from Relativistic and
Quantum Mechanics.

The variable proper mass concept introduced in the present work is not a hypothesis: 
it comes from the Lagrangian formulation of Relativistic Mechanics, and it is 
the alternative to the GRT curved space philosophy. Moreover, it is   
the alternative to the concept of the proper mass constancy in 
current field theories. Let us compare some our results with 
GRT. In the case of a non-rotating spherical symmetric field    
Schwarzschild found the
exact solution to the GRT field equations:
\begin{equation}
ds^2=c_0^2dt^2(1-2r_g/r)^2-dr^2/(1-2r_g/r)^2-r^2d\phi^2
\label{2.11}
\end{equation}
In the weak-field approximations, which is convenient for the purpose of our
comparison, it is equivalent to the form
\begin{equation}
ds^2=c_0^2(dt/\gamma_r)^2-(\gamma_rdr)^2-r^2d\phi^2
\label{2.12}
\end{equation}
where $\gamma_r=1/(1-r_g/r)$. The Schwarzschild solution reflecfts 
GRT metric relations, as follows:
\begin{equation}
dr'=\gamma_rr,\ \ \ dt'=dt/\gamma_r,\ \ \ c'=c_0,\ \ \  m'=m_0
\label{2.13a}
\end{equation}
and it is apparently consistent with gravitational 
experimental data. It should be emphasized that the latter have been obtained 
essentially from ``weak-field'' experiments. One can deduce the coordinate
speed and the arc speed from (\ref{2.11}) by putting the ``light-cone
condition'' $ds^2=0$; the expressions will appear the same as we have in
(\ref{1.29}) and (\ref{1.28}), but they should be regarded as ``metric
induced'' quantities. In the GRT treatment of experiments 
the speed of light $c_0$ and the proper mass
of elementary particle are considered field-independent physical constants. 
This assumption is a part of  
the GRT metric methodology of
operations with ``wristwatches'' and ``rigid measuring rods''. 
It is not possible to establish 
a relationship of those operations with physical
processes of probing a field with test particles, and the role of the 
photon in a metric determination is not clear. Recall
that our metric is
\begin{equation}
dr'=dr,\ \ \ dt'=dt,\ \ \ c'=c_0/\gamma_r,\ \ \ m'=m/\gamma_r
\label{2.13b}
\end{equation}
which has uniquely resulted from the analysis of a photon and particle motion 
in the gravitational field and the methodology of transportable and 
reproducible standard units. While the GRT metric \label{2.12} contains singularities 
under strong-field conditions, 
our metric \label{2.13b}is free of singularities in the whole range of energy. 
Consequences of a metric difference are further illustrated by examples
of comparison. GRT exact predictions are expressed, as
usually, in ``bookkeeper'' coordinate system
($r$,$t$), which is a Minkowski space coordinate system. 
It is easy to transform the GRT results into a local (``curved'')
coordinate system in accordance with (\ref{2.13a}).


\subsubsection{Examples}
\label{section.2.1.4)}
{\it The proper time and energy.}

The proper time determination and related issues in GRT is discussed in
detail in \cite{1}. It is shown  there that a free-floating particle conserves its total
mass-energy: the formula $E=m_0c_0^2$ remains valid in a ``bookkeeper file''
for the whole period of particle flight. Recall that kinetic energy 
does not appear there in any form. The GRT proper time of a free-floating 
particle to be compared with our result (\ref{2.8a}) is
\begin{equation}
d\tau(r)=dt/\gamma_r^2
\label{2.14}
\end{equation}
Note, that our dynamical 
gamma factor is $\gamma_r=1/(1-r_g/r)$ while in
the Schwarzschild metric its exact expression is
$\gamma_r=1/\sqrt{1-2r_g/r}$; this difference is not important as far as we
deal with a weak-field approximation.
\\

{\it The radial speed.} 

According to GRT \cite{1}, the relative
speed of a particle in a radial fall is 
\begin{equation}
\beta=(1-2r_g/r)\sqrt{1-(1-2r_g/r)/\gamma_0^2}
\label{2.16}
\end{equation}
It should be compared with our result (\ref{1.27}), which is valid at $r>r_g\ge R$. 
The equation (\ref{2.16}) shows that from the viewpoint of the observer 
at infinity a particle dropped from rest begins to accelerate, then at some point  
starts decelerating and eventually stops at $r=2r_g$. The bigger initial kinetic 
energy, the greater a "resisting" force arising so that the speed of the particle cannot exceed the 
coordinate speed of light. Strangely enough, if $\gamma_0 \ge\sqrt{1.5}$, 
the particle will never accelerate in a gravitational field. 
In the SRT approach the particle falling onto a gravitational center always 
accelerates; if initial kinetic energy is big enough, 
its speed may exceed the coordinate speed of light (but not the ultimate speed $c_0$).   
\\

{\it The GRT black hole concept.} 

The known GRT
``black-hole'' concept is not a pure GRT result: an additional postulate
of a ``gravitational collapse'' is needed. The process allegedly leads to 
a formation of singularity points in space if a mass
concentration exceeds a critical value. The Schwarzschild 
metric and equations 
(\ref{2.14}-\ref{2.16}) reflects physics of the black hole concept. There is 
a ``horizon condition'', which requires a change of a metric signature 
when the particle crosses the Schwarzschild sphere. In our approach 
the ``black hole'' phenomenon, as presented in GRT, does not exist.
\\

{\it The Einstein's elevator.} 

A freely falling elevator in GRT imaginary experiments is usually
associated with the idea of Equivalence Principle (EP). As is known, the
EP rests on the classical postulate of gravitational and inertial mass
equality, which was formulated in terms of Newtonian Physics. In view of 
relativistic generalization of mass-energy concept presented in this work,
the EP does not have any physical sense. 

The metric concept, as any concept in Physics, is subject to experimental
verification, as discussed further.

\subsection{Classical tests}
\label{section.2.2}

There are four types of classical experimental tests of GRT we have meant to
refer to: the gravitational red-shift, the bending of
light, the time delay of light, and the advance of perihelion of planets.
All of them are related to weak-field conditions and often presented in
popular literature as a solid GRT confirmation. Detailed analysis of
gravitational experiments may be found in \cite{6}. Below we make a brief
review of the tests from different points of view. As was expected, \em there is
no meaningful numerical difference between GRT and SRT-based predictions of weak-field
effects in spite of their different physical interpretation \em.

\subsubsection{The gravitational red-shift}
\label{section.2.2.1}

The term ``red-shift'' means that the wavelength of a photon emitted by an
atomic clock at some point of lower potential appears to be increased when
detected at some point of higher potential. It is impossible from this
type of observations alone to find out what happens to a photon during its
travel from an emitter to a detector.

In our approach, the effect is a result of new concepts of the
proper mass and the photon. Specifically, it is due to the shift of the
emission-detection resonance line: as follows from (\ref{2.1}-\ref{2.4}), 
it is $\delta\lambda/\lambda=\delta m /m=-r_g\delta(1/r)$. 
In GRT, the proper mass is constant; hence, the standard resonance frequency
of an emission-absorption line does not depend on a potential. One might 
think that the wavelength
shift should be due to the change of either the speed of wave propagation 
or the frequency of the photon. In fact, in GRT treatment the effect 
is due to the curved space metric. Sometimes it is treated in geometrical form,
from which it follows that the atomic clock rate depends on ``potential'':
\begin{equation}
\delta t(r)=\delta t_0(1-2r_g/r)^{1/2}\approx\delta t_0(1-r_g/r)
\label{2.13}
\end{equation}
The corresponding ``energetic'' explanation is that the photon is attracted by
the gravitational center and loses its energy while ``climbing up''; 
consequently, the frequency decreases while the speed of light is
constant. However, the ``energetic treatment'' is not relevant 
because the concept of
potential energy is absent in the GRT arsenal. Some relativity experts still argue
that the atomic resonance frequency does depend on the potential. They
conclude that this is a relative binding shift in frequencies of an emitter
and a detector that causes the red-shift, while the speed of light is
constant \cite{7}. However, this would be contradictory to the GRT basic
concept of proper mass constancy.

\subsubsection{The bending of light}
\label{section.2.2.2}

As was discussed, in SRT Mechanics the bending of light is explained by a
dependence of the speed of light on the gravitational potential; that is, due
to the ``gravitational refraction'' rather than the ``gravitational
attraction''. These are different physical phenomena The gravitational
attraction (as opposed to the refraction) changes a photon frequency: this 
does not follow directly from the observations. According to our metric, 
the frequency of a photon in flight is constant, in agreement with the 
conservation of angular momentum. Calculations of the bending effect, 
if conducted on the basis of the angular momentum conservation and in the model 
of gravitational refraction, give the same result. 

\subsubsection{The time delay of light}
\label{section.2.2.3}

The time delay effect was brought to public attention by Shapiro in 1964 \cite{6}), and
it was confirmed in measurements of a radar echo delay for electromagnetic
pulse passing near the Sun. Nether a varying speed of light nor a gravitational
attraction of photon was ever mentioned in the consistent GRT treatment 
of the effect: 
the curved 4-space metric is sufficient for the weak-field evaluations.
(see, for example, \cite{6}). In our approach, the time of flight of
a light signal can be found by integrating over the path with the coordinate speed
(\ref{1.23}). Similar result is obtained in GRT from the photon geodesic equation.   

\subsubsection{Planetary perihelion precession}
\label{section.2.2.4}

This test (unlike the previous ones) is related to a massive 
particle motion in a
gravitational field. In the weak-field approximation, one may consider the classical
problem of the orbit with a varying proper mass in the weak-field approximation.
This is the proper mass dependence on field that causes a change of the
angular frequency proportionally to the squared proper mass, and the radial
one inversely proportional to the proper mass. Thus, the relative angular
shift per revolution is found to be $3r_g/a$, where $r_g$ is a gravitational
radius of the Sun, and $a$ is an ``effective'' orbital radius. The GRT
treatment with the similar result is usually based on the comparison of
rates of radial and tangential motion in an ``effective potential'' \cite{1}.

\subsubsection{Experiments with atomic clocks}
\label{section.2.2.5}

With the development of atomic clock technology and the Global
Positioning System (GPS), it became possible to test a gravitational theory
in experiments under the weak-field conditions \cite{1,8}. This is the case, 
the problem solution can be rigorously found in SRT-based Mechanics. 
The current GPS philosophy is allegedly based on
GRT weak-field approximation. In fact, it contains parts of Newtonian Physics,
SRT, and GRT. In our view, there is no proof of consistency of those parts;
difficulties and controversy arise in the treatment of non-inertial effects
similar to those in the ``twins paradox''. Our comparative GPS
Physics analysis showed meaningful differences in GPS operational parameters
calculated by different methods. Unfortunately, the ultimate positioning 
precision in current GPS 
versions appreciably depends on 
an incessant adjustment of time synchronization by reference signals. 
In our view, an academic research program of further PS testing is needed with
usage of data from additional ``clean'' experiments. Such a program would be beneficial for
a development of the next GPS generation of improved performance and greater
commercial quality.

\subsubsection{``Negative experiments''}
\label{section.2.2.6}

In our approach, gravitational waves, as predicted by GRT, do not exist; we predict 
the negative result of the corresponding experiments. The same conclusion was made concerning the so-called 
``frame dragging'' GRT effect.

\subsection{New test proposal: detection of superluminal particles}
\label{section.2.3}

Our treatment of gravitational experimental data and predictions is based 
on SRT physical concepts different from GRT ones. This conceptual 
difference can be resolved by experiments having a falsifying 
power. We propose an experimental test to check our prediction of 
the existence of superluminal particles in a gravitational field. 
Our results show that light propagates in a gravitational field as in 
a refractive medium; consequently, the speed of a particle can exceed 
the local speed of light. Superluminal particles should be accompanied by a 
specific Cherenkov radiation which could be registered. This prediction may be 
tested under Earth conditions in experiments with
ultrahigh-energy cosmic ray particles. Equating expressions for the particle speed
(\ref{1.27}) and the photon speed (\ref{1.29}) one may find a threshold
energy, at which a particle becomes superluminal. The typical altitude of a
satellite with astrophysical instrumentation is in the range 400-600 km
above the Earth surface; for a corresponding threshold energy 
$E_{th}=\gamma_{th}m_0c_0^2$ the gamma factor is about $\gamma_{th}\approx 2\cdot 10^4$, which gives
a proton $E_{th}=2\cdot 10^{13}$ $\it {eV}$. Therefore, we expect that protons in the
energy range $E>2\cdot 10^{13}$ $\it{eV}$ are superluminal and should emit the 
Cherenkov radiation. A rough estimate of a detectable flux of 
superluminal cosmic ray protons is about one particle per
square meter per day; therefore, the experiment is realistic.

Quite probably, the gravitational Cherenkov radiation from superluminal cosmic ray
protons and, possibly, electrons has been already observed but mistreated. We mean the discovery of the so-called 
Terrestrial Gamma Flashes (TGF) \cite{9} which was speculated to be a new atmospheric phenomenon. The TGF events have rare statistics and were accidentally detected in Burst and Transient Source Experiment (BATSE) in the  NASA program of the Compton Gamma Ray Observatory (CGRO) during 1994-2000 period. Gamma spectrometric detectors were installed aboard the BATSE satellite orbiting at 450 km height with the objective to register gamma ray bursts (GRB) from deep space. The experiment was less suited for registration of gamma flashes seemingly coming from Earth (as should be the case in Cherenkov radiation from superluminal cosmic particles). TGF characteristics occured to be very similar to what one could expected in the proposed test experiment. For example, TGF is typically a gamma burst comprising a low-energy photon train of total energy up to hundred $\ {keV}$; its duration is about 2 $\ {ms}$. Those flashes are randomly beamed upward in a narrow cone, and its sources seem to be uniformly distributed in space. Unfortunately, the BATSE program was terminated, and additional (similar to BATSE) experiments are needed to verify our prediction.

\section{On the problem of unified field theory}
\label{section.3}

\subsection{The unification concept}
\label{section.3.1}

In our view, a thorough conceptual consideration of the problem of a force 
unification is needed prior to deriving field equations; 
some concepts and terms would be reformulated. 
In the following discussions we speculate 
how the introduction of the new mass-energy concept 
can resolve difficulties of current field theories and what are the 
perspectives of a new theory development. The central statement of this work is 
that the proper mass constancy in the metric determination in conventional field
theories leads to inherent contradictions. 
For example, in Relativistic Electrodynamics 
there is the controversial problem of distinguishing between 
material and field parts of the total energy-momentum tensor.
Similarly, there is the GRT problem of a ``modification'' of energy-momentum pseudotensor. 
The field singularity problem, is, of course, the most fundamental. 
For discussions of such issues see, for example,
\cite{4,10,11,12}. Our conclusion was that in order to eliminate 
the root of these problems, a revision of the current mass-energy concept
is needed . A test particle in both GRT and
Electrodynamics and other field theories is actually 
treated as a point limit of an ``absolutely rigid
ball'' with a however small (observed) mass and charge. Consequently, one can think of 
an ideal particle not perturbing a field. However, an absolutely rigid particle 
has to have an infinite ``self-binding'' energy (self-energy).
Its mass equivalent is a ``bare mass'' which should
be compared with the ``observed mass''. The latter appears in 
particle interactions and can be theoretically defined by the operation of
subtracting an infinite self-energy. This is an artificial ``cut-off'' 
(the so-called renormalization) procedure justified as far as results are 
in agreement with experiments. 
In the concept of a renormalizable field theory 
the roles of test particle and photon in probing a field are vaguely specified; 
consequently, the question arises what the terms 
"mass" and "field" meant after all. Our metric analysis showed that those terms aquire 
a clear physical meaning if a theory is formulated in a covariant form with identical symmetries
in the 4-coordinate and 4-momentum complementary space; 
``modifications'' or any arbitrary operations cannot be tolerated. In the consistent 
Lagrangian formulation of Relativistic Mechanics the variable proper mass becomes a field 
in which "the field strength parameter" characterizes the rate of proper mass 
exhaustion. This phenomenon is a physical process resulting in a natural elimination of 
the classical $1/r$ singularity.

The predictive success of GRT and Relativistic Electrodynamics was
impressive but not surprising because both theories were tested and applied
under weak-field conditions (by our criteria $r_g/r\ll 1$, $r_a/r\ll 1$),
while a theory validity under strong-field conditions was never proven. We
assert that both GRT and QED are, in fact, weak-field approximations of a
more general unified theory, discussed further.

We found that both gravitational and electromagnetic field, though having different structure, 
are related to the common source; clearly, this is the clue for the concept of
a unified theory. The electromagnetic field in the Maxwell theory is a vector field described 
by classical potentials:      
\begin{equation}
\phi_e(r)=\frac{1}{4\pi\varepsilon}\int_v\frac{\rho _e(r')}{|r'-r|}dr'^3
\label{3.1}
\end{equation}
\begin{equation}
{\bf A}=\frac{\mu}{4\pi}\int_v\frac{{\bf j}_e(r')}{|r'-r|}dr'^3
\label{3.2}
\end{equation}
while the gravitational field is characterized by a pure scalar
potential $\phi_g(r)$, which is similar to an attractive mode in (\ref{3.1}). 
\em Our idea of field unification is to add the gravitational (mass) source and 
mass-energy (neutral) current to corresponding 
parts of the electromagnetic field in the covariant form. \em 
Consequently, field strength parameters will
be automatically coupled and will play the role of dynamical ``feedback'' variables.
Obviously, the variable proper mass makes the unified field theory  non-linear. One can try to
realize this concept within the Lagrangian formulation by 
applying the variational principle to the extremal proper mass problem. It is expected that
equations of motion will consistently
describe the total energy-momentum tensor of the particle system and yield an
electromagnetic field and a massless mediating boson field. However, it is not 
immediately clear how to formulate the problem of multiple 
interacting sources of unified field. In the linear field concept with 
the proper mass being constant the superposition principle holds, and the $1/r$-potentials 
are usually meant retarded. In our field concept 
a point-like source and a point-like test particle form a system of two interacting 
particles with the radius of interaction (field strength parameter) determined by 
properties of both particles. It seems that retarded and advanced potentials 
are needed to account for both outgoing and ingoing interfering waves.

It would be reasonable to consider a massless boson mediating field in the
Klein-Gordon framework with a variable proper mass. Spin properties of sources  
are not specified there. We assume 
that force transmitting virtual (pure imaginary) photons  
are of two types: longitudinal and scalar (time-like) photons; the latter 
have to contribute to mediating 
a scalar gravitational interaction. Real (transverse) photons are not mediators; 
they have the status of free particles which can exist independent of sources and  
be utilized in cross-section measurements. 
This is in agreement with the Gupta-Bleuler formalism \cite{12a,12b} 
distinguishing between observable
(Hermitian) and mediating (anti-Hermitian) particles. Thus, we want to
interpret it in a ``strong'' form: all interactions are due solely to
anti-Hermitian photons.  
They are physical vacuum excitation states resulting 
from massive particle interactions. In this scheme gravitational properties of real photons 
should be explained in terms of their interaction with the boson field.

In the metric analysis of photon properties we have used quantum-mechanical 
relations characterizing the de Brogli waves.   
This gives us the thought that understanding of the de Brogli phenomenon should 
be an important step towards the development of unified field theory (as discussed next). 

\subsection{The de Broglie waves and the boson field}
\label{section.3.2}

The discovery of particle waves was made by Louis de Broglie in 1923-1924.
One may think that the de Brogli waves are described by the Shroedinger
equation. However, it is not true because this phenomenon is essentially
relativistic. De Broglie was never satisfied with probabilistic (``Copenhagen
School'') interpretation of Quantum Mechanics, and since 1924 he kept working on
his own idea of a next-level (``double-solution'') theory \cite{13}. The
physical meaning of the Shroedinger $\psi$-function as a probability
amplitude of particle waves has been debated for decades, the nature of the
De Broglie waves and so-called ``entangled states'' being the central issue
(see, for example, \cite{14,15,16}). In our view, the question about causality
versus probability interpretation of microscopic experiments is ill-posed, while the 
nature of the de Brogli waves is a real problem. 
Our arguments are as follows.

The de Brogli waves are commonly known from observations of free moving
particles, first of all, in interference experiments. Let us consider the dynamics of
the wave origination. In the familiar example of
an attractive interaction, the following equation describes the
dynamical energy conservation balance:
\begin{equation}
p^2/c_0^2+m^2=m_0^2
\label{3.3}
\end{equation}
with $m/m_0=(1-x_s/x)\le1$ where $x_s$ is a field strength parameter.
The de Broglie wavelength $\lambda_{dB}$ due to the momentum transfer may be
derived from the Einstein's and de Broglie's quantum-mechanical expressions
$p=h/\lambda_{dB}$ and $m_0=hf_0/c_0^2$ together with our formula for the
particle momentum $p=m_0c_0\beta$. Then, the equation (\ref{3.3}) in terms of
frequencies is equivalent to the following:
\begin{equation}
f_{dB}^2+f^2=f_0^2
\label{3.4}
\end{equation}
\begin{equation}
(\lambda_0/\lambda_{dB})^2+(m/m_0)^2=1
\label{3.5}
\end{equation}
The field-dependent frequency $f(t)$ is
related to the proper mass of the atomic clock. The corresponding proper wavelength of the 
atomic clock oscillation 
is $\lambda(t)=c_0/f(t)=h/m(t)c_0$, at initial moment being equal to
$\lambda_0=h/m_0c_0$. According to our metric relations (\ref{2.1}), it
increases during particle acceleration. Note that the de Broglie waves are 
determined by the spatial part of the 4-wave
vector, while the proper mass component is 
a time-like quantity. Thus, the de Brogli wave characteristics
are momentum (space-like) quantities. Unlike the proper wavelength, 
the de Brogli wavelength decreases
during particle acceleration in free fall. The waves should have the form
$a(x,t)\propto\cos(ft-kx))$, where $f(t)=c_0^2m(t)/h$, $k(x)=c_0m_0\beta(x)/h$,
$\beta(x)=\sqrt{1-(m/m_0)^2}$, $m/m_0=(1-x_s/x)$, $x=x(t)$.
Thus, a connection is seen between conservative field symmetries previously discussed 
and the de Broglie waves. Characteristics of the waves are present in 
relativistic metric relations, which we derived from completely different principles. 
The whole picture can be
visualized. Imagine a particle at rest producing a static
spherical symmetric field. It has one degree of freedom (the scalar mode)
what may be thought of as the sphere pulsation. The pulsating sphere
in motion exhibits known relativistic effects of length contraction and time
dilation giving rise to the vector mode of oscillation that is, the de Broglie wave
phenomenon. The group speed is the particle speed while the corresponding 
phase speed exceeds the ultimate speed of light. 
In traditional interference experiments the wave becomes polarized in 
a plane after coming through a single slit. An experimentalist can rotate the 
plane of polarization by creating an accelerating field between the slit and a screen. 
The plane rotation is the predicted 
relativistic effect which could be verified, in principle.

It is seen that the ``probability wave'' concept in the
non-relativistic quantum-mechanical theory is an approximation, which could be somehow
justified under weak-field conditions (a proper mass constancy). 
In our relativistic picture, the source of the de Brogli waves is the moving particle 
in a fixed reference frame
(the $cp$ term in the equation $cp=hf_{dB}$). Thus, a particle interference pattern 
is frame dependent. At the same time, the de Brogli waves are associated 
with an excitation of physical vacuum states (virtual photons) in the process of transforming 
a proper mass into a kinetic one. The excitation process  
propagate in space with the ultimate speed of light, and it should be treated 
in terms of coherent ingoing and outgoing waves of the boson field.
From this point of view, the wave nature of particles 
(the de Brogli waves, tunneling effects and entangled collective states) may be understood; however, 
the term  ``physical vacuum'' needs to be clarified.   
Intuitively, one may think of a spherical shell of universe matter 
as the source of ``physical vacuum field'' with a finite energy density. Then, the
physical vacuum in the shell is a background field for local fields in a cosmological potential well. 
In this sense, to reach ``infinity'' means to get rid of all local fields. For an atomic 
electron it could be a fraction of millimeter away. A single free 
particle moving in the space of a constant residual potential is in equilibrium with
the physical vacuum (the universe): there is no net mass-energy, or a virtual photon, current
between the particle and the universe. This is the particle-particle interaction, which breaks this equilibrium 
and results in the net current between interacting particles and the universe.  
The direction of current is determined by the field gradient depending on the type of interaction. 
The process is traditionally 
described in the quantum-mechanical concept of the photon exchange mechanism 
to be revisited in the future non-linear theory. We emphasize here the importance 
of relativistic mass-energy concept (a proper mass variability) with cosmological connections as well as 
a possible universal role a boson field (virtual photons) in a unified field theory. In fact, we assume 
that virtual photons mediating gravitational and electromagnetic forces are revealed in 
a form of the de Brogli waves which should be subject to further experimentation.  

\subsection{On the spinor field concept}
\label{section.3.3}

A consideration of particle spin 
properties will require a next-level theoretical concept, presumably,
in the framework of the spinor (Dirac) field. Again, the concept of variable proper mass 
should be incorporated in the theory. Then, the boson field 
could be treated as a composite field. A spin is both 
quantum-mechanical and relativistic quantity 
and, as such, seems to be poorly understood. 
A massive fermion spin, which appears in a helicity operator of a massive particle 
$(\bf{\sigma}\cdot{\bf p})$, is not Lorentz invariant; 
but it is invariant in the case of a 
massless Dirac neutrino. Obviously, this is the absence of the proper mass that makes a 
neutrino spin Lorentz invariant. In classical terms, 
one can associate the spin with the particle rotation. 
But what can give a neutrino a rotational mode when there is no 
neither proper mass nor magnetic moment? We think that our teatment 
of the de Brogli waves allows us to gain an insight into the 
nature of spin. As was shown, a massive particle is a relativistic object 
characterized by the 4-wave vector having a spatial (the de Brogli wave) 
and a time-like (the scalar wave) parts. The corresponding sources of 
virtual photons are the momentum of a massive particle. Recall that the total energy of 
any particle is a magnitude of its 4-momentum. In a process of proper-to-kinetic mass 
transformation, real (transverse) photons can emerge as the result of the proper mass 
annihilation or electromagnetic transitions between resonance states in a bound systems
(for example, in atoms). Note that such systems are formed by attractive forces. Thus, the photon 
having no proper mass may be considered a kinetic mass-energy quantum, that is the quantum 
of a vector field, or electromagnetic energy carrier. In this sense, the photon is a pure 
space-like real object. Could a pure time-like real photon exist as a proper mass quantum?
There are some arguments supporting this hypothesis, given in brief below. 
\\

{\it The neutrino is a time-like massless real fermion.}
We assume that a massive particle is a physical vacuum resonance in a cosmological 
gravitational field. Let us consider the particle to be a droplet of a Bose-condensate with 
a proper mass quantity proportional to 
the number of condensed photons. Hypothetically, an elementary fermion, like an electron or a proton, 
is a stable ground-state formation finished with a valence proper mass fermionic quantum and 
``entangled'' through it with universe matter. 
The neutrino is a candidate for this quantum. 
We assume that neutrinos are produced in reactions with a special repulsive force, 
the so-called degenerate-pressure force. The latter arises, for example, 
between a proton and an electron at distances smaller then the Bohr radius. 
In accordance with our mass-energy concept, the electron proper mass under such conditions 
becomes greater than that at infinity. When it reaches the 
resonance state about 106 Mev, the muon is created, which could be 
absorbed by the proton, a neutron being formed. In this picture, the 
muon should be considered an exited electron being subject to decay with 
neutrino-antineutrino emission. When the muon is absorbed by the proton, the electron undergoes a stage of 
coupling to antineutrino with the neutrino flying away. If true, the electron-antineutrino bound 
state is real that is, the electron can exist in a free baryon state 
(as a baryonic electron), which could be detected. A neutron decay is accompanied by the 
bosonic electron decay. Therefore, the neutrino plays the role of the proper mass quantum able 
to couple in pairs (forming virtual photons) or to any fermionic particle (as in the bosonic electron). 
There is a principle difference between the real photon and the neutrino. The photon carries the momentum 
but does not have an angular momentum: in a circular polarization mode it reveals a rotation of 
a plane polarization due to phase shift with no rotational kinetic energy. 
Unlke the poton, the neutrino being a proper mass quantum carries inner 
angular momentum with no linear momentum. Such a particle has one degree of freedom and 
should be called a pure time-like particle. Its behavior in a gravitational field should be investigated. 
\\

{\it Electroweak model?} Reactions in which neutrinos are involved are called ``weak interactions''. 
In our approach they do not manifest special fundamental forces to be unified with the electric ones. 
Therefore, the so-called electroweak model of spontaneously broken symmetry \cite {weak} is not needed.  
\\

{\it There is no violation of P or T symmetry: strictly speaking, such symmetries do not exist in a relativistic world.} 
Considerations of the parity issue gives us the thught that there is no physical reason for the mirror symmetry in reactions; ``the neutrino'' and ``the antineutrino'' automatically have opposite 
handedness by virtue of angular momentum conservation, and the question of 
existence of both right-handed and left-handed neutrinos 
(or the world being left-right symmetric) becomes ill-posed. The parity is a concept of 
non-relativistic (``probability wave'') quantum mechanics. We assert that the 
CPT theorem should be replaced by a relativistic invariance concept of the 4-coordinate reverse symmetry 
with a properly defined conjugation operator in a boson or spinor field. In other words, the concept of antimatter needs to be reconsidered.  
\\

{\it The mediating boson field is composed of virtual neutrino pair states.}
In the Dirac framework virtual photons may be considered 
virtual neutrino pairs being ``force transmitters'' (but not energy carriers) 
in gravitational and electromagnetic interactions, ``weak'' reactions included. This would provide a 
natural connection between unified theories in the Klein-Gordon 
and Dirac framework. Problems of Particle Physics could not be formulated 
in the above sketch: a higher-rank theory should be speculated. Probably, one should come back to Heisenberg's 
ideas of a theory of matter \cite{18,19,20}. 
To our knowledge, his goal was to explain a particle structure and, probably, the nature of mass, in the 
concept of fundamental spinor $\psi(x)$ being a
self-interacting field operator in the non-linear equation
with usual Dirac $\gamma$-matrices:
\begin{equation}
\gamma^\mu\partial_\mu\psi(x)-l^2F[\psi(\tilde\psi\psi)]=0
\label{4.1}
\end{equation}
In general, functional $F[\psi(\tilde\psi\psi)]$ can be constructed from
multiple combinations of $\gamma^\mu$, $\gamma^5$, $\psi$,
$\tilde\psi$. The equation (\ref{4.1}) could be presented as an infinite set
of differential equations corresponding to a multi-point propagator. The latter should give rise to
composite field objects: particles have to arise as non-linear resonance states
(probably, of soliton-type). However, it is not clear how this approach could be 
practically realized and narrowed
to solving the gravitational problem, which is the main issue of the present work. 
In our view, a future physical theory of matter should be essentially a cosmological theory.
\\

{\it Does the neutrino play a fundumental role in the structure of matter?}
The electromagnetic field reveals quantum-mechanical effects mostly on a scale 
of atomic structures. At the same time, gravitational forces become dominant 
in the astronomical and cosmological range with a continuous energy spectrum 
when the de Brogli waves do not play any role. 
One may reckon that in current practical applications of the gravitational theory 
Quantum Mechanics is not necessary: the two fields seem to occupy 
separate niches. Then, why does one need to unify them? 
There is the general opinion among the physical community  that the problem of field 
unification and quantization is important under extreme astrophsical 
conditions, and it is fundamentally important for the progress of our 
understanding of the structure of matter at the ``ultimate bricks'' level. 
We believe that the neutrino should be put in the ``bricks'' list, 
and our attempt to unveil the neutrino mystery is a step in a right direction.             
\section {Conclusion}
\label {Section 4}
The problem of $1/r$ field singularities was studied starting with the question of 
dependence of basic (space-time and proper mass) units and the speed of light on 
the gravitational potential in the Lagrangian formulation of Relativistic Mechanics. 
The criteria of our metric determination were consistency with the covariant form 
of equations of motion in a conservative force field and the agreement with experimental data. 
It was found that the proper mass and the speed of light are field dependent; 
they uniquely characterize the gravitational field. Consequently, the new metric 
provides a unique mapping of the field with the use of a light signal and a test particle 
while experimental data and predictions are consistently treated 
in terms of Relativistic Mechanics. Among the important results are, as follows:
\begin{itemize}
\item
There is no reason for the exclusion of gravitational forces from the theory. 

\item
A field due to gravitational or electric sources is free of singularities; 
therefore, a renormalization procedure is not needed. 

\item
The above approach to the gravitational field problem is consistent with observations 
and is falsifiable. One of the predictions is an existence of superluminal particles in 
a gravitational field. A corresponding experimental test is proposed.

\item
The final conclusion was that a development of a unified field theory is possible 
in which the neutrino is expected to play an important role of a proper mass quantum.           
\end{itemize}

\end{document}